\documentclass[11pt,a4paper]{article}

\usepackage{amssymb}
\usepackage{graphicx}
\usepackage{bm}

\usepackage{mathrsfs}
\usepackage{cite}

\newcommand*\xbar[1]{%
  \hbox{%
    \vbox{%
      \hrule height 0.5pt 
      \kern0.5ex
      \hbox{%
        \kern-0.1em
        \ensuremath{#1}%
        \kern-0.1em
      }%
    }%
  }%
}

\begin{document}

\title{Relation between leading divergences in nonrenormalizable $4D$ supersymmetric theories}

\author{
Ali Lakhal \\
{\small{\em Aix-Marseille University, Faculty of Science, 13331 Marseille, France}}\\
\\
and Konstantin Stepanyantz\\
{\small{\em Moscow State University}}, {\small{\em  Faculty of Physics, Department  of Theoretical Physics}}\\
{\small{\em 119991, Moscow, Russia}}\\
}

\maketitle

\begin{abstract}
We consider an ${\cal N}=1$ nonrenormalizable supersymmetric gauge theory with the superpotential quartic in the chiral matter superfields. With the help of the Slavnov's higher covariant derivative regularization it is demonstrated that (in the lowest nontrivial order) the leading power divergent quantum correction to the gauge coupling constant is given by an integral of double total derivatives with respect to the loop momenta. The result obtained after calculating this integral turned out to be proportional to the corresponding quantum correction to the kinetic term of the matter superfields. More exactly, in the considered approximation the quadratically divergent contributions to the gauge coupling and to the kinetic term of the chiral matter superfields are related by an equation analogous to the exact NSVZ $\beta$-function for the renormalizable case.
\end{abstract}

\section{Introduction}
\hspace*{\parindent}

Supersymmetric theories are of considerable interest both for particle physics and theoretical research in quantum field theory. For instance, the renormalization group running of gauge couplings in supersymmetric extensions of the Standard Model compatible with predictions of Grand Unified Theories is a strong evidence in favor of supersymmetry \cite{Ellis:1990wk,Amaldi:1991cn,Langacker:1991an}. Moreover, the absence of quadratically divergent quantum corrections to the Higgs boson mass in the supersymmetric case allows for avoiding its fine tuning at the unification scale, see, e.g., \cite{Mohapatra:1986uf,West:1990tg} and references therein. This fact occurs because at the quantum level supersymmetric theories have much better behaviour than their nonsupersymmetric analogs \cite{Gates:1983nr,West:1990tg,Buchbinder:1998qv}. In particular, divergences are absent in ${\cal N}=4$ supersymmetric Yang-Mils theory in all orders \cite{Sohnius:1981sn,Grisaru:1982zh,Howe:1983sr,Mandelstam:1982cb,Brink:1982pd} or exist only in the one-loop approximation in ${\cal N}=2$ supersymmetric gauge theories \cite{Grisaru:1982zh, Howe:1983sr,Buchbinder:1997ib}. (This in particular implies that one can construct finite ${\cal N}=2$ theories \cite{Howe:1983wj} with the help of such a choice of a gauge group and a representation for the chiral matter superfields that the one-loop $\beta$-function vanishes \cite{Banks:1981nn}.) Essential cancellations of ultraviolet divergences occur even in theories with unextended (${\cal N}=1$) supersymmetry, which are especially interesting for phenomenology. According to \cite{Grisaru:1979wc}, in this case the superpotential

\begin{equation}
W = \frac{1}{2} m_0^{ij} \phi_i\phi_j + \frac{1}{3} \lambda_0^{ijk} \phi_i\phi_j\phi_k
\end{equation}

\noindent
does not receive divergent quantum corrections, which in turn leads to certain equations  relating renormalization of masses and Yukawa coupling to the renormalization of the chiral matter superfields,

\begin{equation}
m^{ij} = m_0^{mn} (\sqrt{Z})_m{}^i (\sqrt{Z})_n{}^j;\qquad \lambda^{ijk} = \lambda_0^{mnp} (\sqrt{Z})_m{}^i (\sqrt{Z})_n{}^j (\sqrt{Z})_p{}^k,
\end{equation}

\noindent
where the subscript $0$ marks bare values and the renormalization constant for the chiral matter superfields is determined by the equation $\phi_i = (\sqrt{Z})_i{}^j \phi_{j,R}$.

A much less obvious fact is that the renormalization of the gauge couplings in ${\cal N}=1$ supersymmetric theories is also related to the renormalization of the chiral matter superfields by a certain relation called the NSVZ equation \cite{Novikov:1983uc,Jones:1983ip,Novikov:1985rd,Shifman:1986zi},\footnote{Eq. (\ref{NSVZ_Equation}) is written for a theory with a single gauge coupling. The generalization to theories with multiple gauge couplings (which, in particular, include MSSM) are also known \cite{Shifman:1996iy,Korneev:2021zdz}.}

\begin{equation}\label{NSVZ_Equation}
\frac{\beta(\alpha,\lambda)}{\alpha^2} = - \frac{3C_2-T(R) + C(R)_j{}^i \gamma_i{}^j(\alpha,\lambda)/r}{2\pi(1-\alpha C_2/2\pi)},
\end{equation}

\noindent
where the renormalization group functions (RGFs) are defined by

\begin{equation}\label{RGFs_Renormalized}
\beta(\alpha,\lambda) \equiv \frac{d\alpha}{d\ln\mu} \bigg|_{\alpha_0,\lambda_0=\mbox{\scriptsize const}};\qquad
\gamma_i{}^j(\alpha,\lambda) \equiv \frac{d\ln Z_i{}^j}{d\ln\mu} \bigg|_{\alpha_0,\lambda_0=\mbox{\scriptsize const}},
\end{equation}

\noindent
$\alpha=e^2/4\pi$ is the gauge coupling constant, and the derivatives with respect to the renormalization point $\mu$ should be taken at fixed values of the bare couplings $\alpha_0$ and $\lambda_0$. In our notation, $r$ is the dimension of the gauge group $G$. In what follows we will always assume that the generators $t^A$ of its fundamental representation satisfy the normalization condition and the commutation relations given respectively by the equations

\begin{equation}
\mbox{tr}(t^A t^B) = \frac{1}{2}\delta^{AB};\qquad [t^A, t^B] = if^{ABC} t^C,
\end{equation}

\noindent
where $f^{ABC}$ are totally antisymmetric structure constants of the group $G$. For an arbitrary representation $R$ the generators denoted by $T^A$ satisfy the same commutation relations $[T^A, T^B]= if^{ABC} T^C$. Using them, we define the group Casimirs by the equations

\begin{equation}
\mbox{tr}(T^A T^B) \equiv T(R)\delta^{AB};\qquad f^{ACD} f^{BCD} \equiv C_2 \delta^{AB};\qquad (T^A T^A)_i{}^j\equiv C(R)_i{}^j.
\end{equation}

The NSVZ equation (\ref{NSVZ_Equation}) establishes the relation between the gauge $\beta$-function in a certain approximation and the anomalous dimension of the matter superfields in all previous loops. However, multiloop calculations made in \cite{Jack:1996vg,Jack:1996cn,Jack:1998uj,Harlander:2006xq} (see \cite{Mihaila:2013wma} for a review) demonstrated that the NSVZ equation is valid only for certain renormalization prescriptions called ``NSVZ schemes''. In particular, for the most popular in the supersymmetric case $\overline{\mbox{DR}}$ scheme (when a theory is regularized by dimensional reduction \cite{Siegel:1979wq} and the renormalization is made with the help of modified minimal subtraction \cite{Bardeen:1978yd}) the NSVZ equation does not hold starting from the order $O(\alpha^2,\alpha\lambda^2,\lambda^4)$, where the dependence of RGFs on the renormalization prescription becomes important. The general equations describing the scheme dependence of the NSVZ equation can be found in \cite{Kutasov:2004xu,Kataev:2014gxa}.

The all-loop NSVZ renormalization scheme was obtained with the help of the Slavnov's higher  covariant derivative regularization \cite{Slavnov:1971aw,Slavnov:1972sq}. In this case the regularization is made by adding to the action some terms containing higher derivatives which provide the rapid fall of the propagators at large momenta. This allows removing all divergences beyond the one-loop approximation. For regularizing the residual one-loop divergences one should insert some special Pauli--Villars determinants into the generating functional \cite{Slavnov:1977zf}. Although this procedure leads to certain technical difficulties, it has some advantages, especially in the supersymmetric case. For instance, this regularization is formulated in four dimensions and, therefore, has no problems with $\gamma_5$ definition. For supersymmetric theories it can be constructed in a manifestly supersymmetric way with the help of the superfield technique \cite{Krivoshchekov:1978xg,West:1985jx}. Moreover, for supersymmetric theories the higher covariant derivative regularization allows for revealing some properties of quantum corrections that cannot be detected when using the dimensional technique \cite{Stepanyantz:2019lyo}. For instance, the calculations made with this regularization
\cite{Smilga:2004zr,Pimenov:2009hv,Buchbinder:2014wra,Buchbinder:2015eva,Aleshin:2016yvj,Shakhmanov:2017soc,Shakhmanov:2017wji,Kazantsev:2018nbl} demonstrated that the $\beta$-function of supersymmetric theories is given by integrals of double total derivatives with respect to the Euclidean loop momenta. Subsequently, this fact has been proved in all orders in \cite{Stepanyantz:2011jy} (for the Abelian case) and \cite{Stepanyantz:2019ihw} (for general renormalizable supersymmetric gauge theories). Such structure allows for calculating one of the loop integrals and reducing a number of integrations by 1. This in turn leads to the relation between the $\beta$-function and the anomalous dimensions of the quantum gauge superfield, Faddeev--Popov ghosts, and chiral matter superfields. This relation is equivalent to the NSVZ equation (\ref{NSVZ_Equation}) if one takes into account the nonrenormalization of the triple gauge-ghost vertices proved in \cite{Stepanyantz:2016gtk}. Thus, it becomes possible to derive the NSVZ equation by direct summing of the perturbative series \cite{Stepanyantz:2020uke} and, as a byproduct, construct a renormalization prescription for which the NSVZ $\beta$-function is valid in all orders. It appeared that if the higher covariant derivatives are used for regularization, then the NSVZ equation is satisfied in all orders for RGFs defined in terms of the bare couplings

\begin{equation}\label{RGFs_Bare}
\beta_0(\alpha_0,\lambda_0) \equiv \frac{d\alpha_0}{d\ln\Lambda} \bigg|_{\alpha,\lambda=\mbox{\scriptsize const}};\qquad
(\gamma_{0})_i{}^j(\alpha_0,\lambda_0) \equiv -\frac{d\ln Z_i{}^j}{d\ln\Lambda} \bigg|_{\alpha,\lambda=\mbox{\scriptsize const}},
\end{equation}

\noindent
where $\Lambda$ is the dimensionful regularization parameter.\footnote{For dimensional reduction the analogous statement is not valid, see \cite{Aleshin:2016rrr} for detail.} Note that, for a fixed regularization, these RGFs depend on the regularization, but do not depend on the way divergences are removed \cite{Kataev:2013eta}, unlike the standard RGFs (\ref{RGFs_Renormalized}). The RGFs (\ref{RGFs_Renormalized}) defined in terms of the renormalized couplings depend on both the regularization and the renormalization prescription. The all-loop NSVZ scheme is obtained for them when the higher covariant derivative regularization is supplemented by minimal subtraction of logarithms \cite{Kataev:2013eta,Shakhmanov:2017wji}. (By definition, in this case the renormalization constants include only powers of $\ln\Lambda/\mu$.)

The knowledge of the renormalization prescription in which Eq. (\ref{NSVZ_Equation}) is satisfied in all orders allows to simplify certain multiloop calculations to a great extent. For instance, using the NSVZ equation the three-loop $\beta$-functions have been obtained for general supersymmetric gauge theories \cite{Kazantsev:2020kfl,Haneychuk:2025ejd} and, in particular, for ${\cal N}=1$ SQCD+SQED \cite{Haneychuk:2025ehb} and MSSM \cite{Haneychuk:2022qvu}. Using this method the four-loop $\beta$-function of ${\cal N}=1$ SQED was calculated in \cite{Shirokov:2022jyd}. The correctness of the result was subsequently confirmed by the direct calculation of supergraphs made in \cite{Shirokov:2023jya}.

However, various supersymmetric models constructed for describing physics beyond the Standard Model (see, e.g., \cite{Raby:2017ucc}) are not renormalizable. Very often their superpotential includes terms that are more than cubic in chiral matter superfields. Because the corresponding couplings have the dimension of mass in a negative power, the presence of such terms breaks renormalizability. (Nevertheless, they can arise effectively in the renormalizable theories after integrating out some massive modes, see, e.g. \cite{Veltman:1980mj}, or are produced by certain supergravity theories \cite{Nilles:1983ge,Lahanas:1986uc}.) As an example of a term quartic in the chiral matter superfields we can present the $SU(5)$ interaction

\begin{equation}\label{Quartic_Term}
M^{-1}\, (5\times\,\xbar{10}\,\times 5_h\times 75_H) =M^{-1}\, 5_i \,\xbar{10}\,{}^{kl} h_j H^{ij}_{kl},
\end{equation}

\noindent
which is sometimes used for explaining some features of the elementary particle mass spectrum \cite{Altarelli:2000fu}. In this case the $SU(5)$ symmetry is spontaneously broken down to $SU(3)\times SU(2)\times U(1)$ by the vacuum expectation value of the Higgs superfield $H^{ij}_{kl}$ in the representation $75$, where the indices range from 1 to 5. Due to the antisymmetry and tracelessness conditions, this vacuum expectation value can be written as

\begin{equation}
H^{\alpha\beta}_{\gamma\delta} = \big(\delta^\alpha_\gamma \delta^\beta_\delta - \delta^\alpha_\delta \delta^\beta_\gamma\big)\, V;\qquad H^{a\alpha}_{c\delta} = - \frac{1}{3}\delta^a_c\delta^\alpha_\delta\, V;\qquad
H^{ab}_{cd} = \frac{1}{3}\big(\delta^a_c \delta^b_d - \delta^a_d \delta^b_c\big)\, V,
\end{equation}

\noindent
where the indices $a,b,\ldots$ range from 1 to 3 and the indices $\alpha,\beta,\ldots$ range from 4 to 5. Then, (taking the Higgs doublet from the superfield $h$ in the fundamental representation of the group $SU(5)$) at low energies the term (\ref{Quartic_Term}) leads to the equation $Y_E = -3Y_D$ for the corresponding Yukawa couplings. This is precisely the relation between the Yukawa couplings of the second generation in the Georgi--Jarlskog textures for the Yukawa matrices\cite{Georgi:1979df} needed for explaining the value of the expression $m_e m_s/(m_\mu m_d)\sim 1/9$.

Therefore, it would be interesting to investigate the properties of ultraviolet divergences in theories in which the superpotential is more than cubic in the chiral matter superfields and, in particular, to reveal whether it is possible to construct an analog of the NSVZ equation in this case. Note that some aspects of ultraviolet behaviour in nonrenormalizable theories (including the supersymmetric ones) were previously discussed, e.g., in \cite{Borlakov:2016mwp,Kazakov:2019wce,Kazakov:2020kbj,Bork:2021mmm,Kazakov:2022pkc,Kazakov:2023raj} and other literature. However, here we are mostly focused on finding properties similar to the NSVZ relation in the renormalizable case. Namely, in this paper we will consider a theory with a superpotential quartic in the chiral matter superfields and investigate whether it is possible to find certain relations between various divergences, analogous to the case of a renormalizable theory.

The paper is organized as follows. In Sect. \ref{Section_Theory} we formulate the theory under consideration and regularize it by higher derivatives. After that, in Sect. \ref{Section_Quantum_Corrections} the lowest (quadratically divergent) quantum corrections to the gauge coupling and to the kinetic term of the chiral matter superfields proportional to the squared quartic Yukawa couplings are calculated. In particular, it is demonstrated that with the higher covariant derivative regularization the leading quantum correction to the gauge coupling is determined by an integral of double total derivative in the momentum space and is related to the leading contributions to the kinetic term of the chiral matter superfields by an equation analogous to Eq. (\ref{NSVZ_Equation}). The alternative way of deriving the integrals of double total derivatives based on the results of \cite{Stepanyantz:2019ihw} is considered in Sect. \ref{Section_Vacuum}, where we argue that some methods for making calculations in the renormalizable supersymmetric theories are also applicable for the nonrenormalizable case. A brief summary of the results is given in Conclusion. Some details and auxiliary calculations are described in Appendices.

\section{${\cal N}=1$ supersymmetric theories with the superpotential quartic in the chiral matter superfields}
\hspace*{\parindent}\label{Section_Theory}

We will consider an ${\cal N}=1$ supersymmetric theory with a simple gauge group $G$ containing the operator of the dimension 4 in the superpotential.\footnote{A similar model without the gauge superfield was investigated in \cite{Bork:2021mmm}.} This model is described by the superfield action

\begin{eqnarray}\label{Action}
&& S = \frac{1}{2 e_0^2} \mbox{Re}\,\mbox{tr} \int d^4x\,d^2\theta\,W^a W_a + \frac{1}{4} \int d^4x\,d^4\theta\,\phi^{*i} (e^{2V})_i{}^j \phi_j \nonumber\\
&&\qquad\qquad\qquad\qquad\qquad\qquad\qquad + \Big(\frac{1}{24}\int d^4x\,d^2\theta\,\lambda_0^{ijkl}\phi_i\phi_j\phi_k\phi_l +\mbox{c.c.}\Big),\qquad
\end{eqnarray}

\noindent
where $\phi_i$ are chiral matter superfields in the representation $R$, and $V$ is the Hermitian gauge superfield, which strength is given by the chiral superfield $W_a \equiv \bar D^2(e^{-2V}D_a e^{2V})/8$. The gauge coupling is denoted by $e_0$, while $\lambda_0^{ijkl}$ are the analogs of the Yukawa couplings. The gauge invariant theory is obtained if these couplings satisfy the constraint

\begin{equation}\label{Yukawa_Invariance}
0 = \lambda_0^{mjkl} (T^A)_m{}^i + \lambda_0^{imkl} (T^A)_m{}^j + \lambda_0^{ijml} (T^A)_m{}^k + \lambda_0^{ijkm} (T^A)_m{}^l,
\end{equation}

\noindent
which we will always assume to be satisfied. The couplings $\lambda_0^{ijkl}$ have the dimension $m^{-1}$, so that the theory is not renormalizable. More exactly, the degree of divergence is given by the expression

\begin{equation}\label{Degree_Of_Divegence}
\omega = 2-n_\phi+n_\lambda,
\end{equation}

\noindent
where $n_\phi$ is a number of external lines corresponding to the chiral matter superfields, while $n_\lambda$ is a number of vertices containing $\lambda_0^{ijkl}$ or $\lambda^*_{0ijkl}$.

Note that the quartic term(s) in the superpotential may appear, for instance, after integrating out certain massive superfields. In this case their values are suppressed by the inverse mass of these superfields, which may be of the order of the unification scale or even of the Planck mass. That is why, despite the presence of power divergences, quantum corrections containing $\lambda_0$ do not become too large when evolving to the scale of unification \cite{Veltman:1980mj}.

To investigate power divergences, we will use the higher covariant derivative regularization formulated in ${\cal N}=1$ superspace. This, in particular, implies that supersymmetry will be a manifest symmetry at all steps of calculating quantum corrections. Moreover, this regularization allows avoiding problems coming from the noninteger space-time dimension, see, e.g., \cite{Jack:1998exa,Gnendiger:2017pys}.\footnote{Note that power divergences can also be investigated with the dimensional technique, see, e.g., \cite{Jack:1990pz,Al-Sarhi:1990nmv,Al-sarhi:1991gdi} for detail.}

Slavnov's regularization is introduced by replacing the original action with its modification

\begin{eqnarray}\label{Action_Regularized}
&& S_{\mbox{\scriptsize reg}} = \frac{1}{2 e_0^2} \mbox{Re}\,\mbox{tr} \int d^4x\,d^2\theta\,W^a \Big[e^{-2V} R\Big(-\frac{\xbar{\nabla}{}^2 \nabla^2}{16\Lambda^2}\Big) e^{2V}\Big]_{\mbox{\scriptsize Adj}} W_a\nonumber\\
&& + \frac{1}{4} \int d^4x\,d^4\theta\,\phi^{*i} \Big[F\Big(-\frac{\xbar{\nabla}{}^2 \nabla^2}{16\Lambda^2}\Big)e^{2V}\Big]_i{}^j \phi_j
+ \Big(\frac{1}{24}\int d^4x\,d^2\theta\,\lambda_0^{ijkl}\phi_i\phi_j\phi_k\phi_l +\mbox{c.c.}\Big),\qquad
\end{eqnarray}

\noindent
containing the higher covariant derivatives

\begin{equation}
\nabla_a \equiv D_a;\qquad \xbar{\nabla}_{\dot a} \equiv e^{2V} \bar D_{\dot a} e^{-2V}
\end{equation}

\noindent
inside the regulator functions $R$ and $F$. In the simplest case, these functions can be chosen in the form $R(x)=1+x^n$, $F(x)=1+x^m$, where $n$ and $m$ are some positive integers, see, e.g., \cite{Stepanyantz:2011cpt}. However, in general, we will only require that they should rapidly increase at infinity and satisfy the conditions $R(0)=F(0)=1$. For instance, it is possible to use the exponential function \cite{Singh:2025fgo}. The regularization parameter $\Lambda$ which is present in the higher derivative terms has the dimension of mass. The subscript Adj in the gauge part of the regularized action (\ref{Action_Regularized}) means that all products should be replaced with commutators,

\begin{equation}
\Big[f_0 + f_1 V + f_2 V^2 + \ldots\Big]_{\mbox{\scriptsize Adj}} X \equiv f_0 X + f_1 [V,X] + f_2 [V,[V,X]] + \ldots
\end{equation}

\noindent
(This in particular implies that $(e^{2V})_{\mbox{\scriptsize Adj}} X = e^{2V} X e^{-2V}$.)

For quantizing the theory, it is convenient to involve the background field method \cite{DeWitt:1964mxt,Arefeva:1974jv,Abbott:1980hw,Abbott:1981ke} in the superfield formulation \cite{Gates:1983nr,West:1990tg}, because it allows constructing the manifestly gauge invariant effective action. Certainly, in order not to lose this main advantage of the background field method, it is necessary to use a manifestly background-invariant gauge fixing term. It is convenient to choose it in the form

\begin{equation}
S_{\mbox{\scriptsize gf}} = - \frac{1}{16\xi_0 e_0^2}\, \mbox{tr} \int d^4x\,d^4\theta\,  \bm{\nabla}^2 v  R\Big(-\frac{\bm{\xbar\nabla}{}^2 \bm{\nabla}^2}{16\Lambda^2}\Big)_{\mbox{\scriptsize Adj}} \bm{\xbar\nabla}{}^2 v,
\end{equation}

\noindent
where, for simplicity, we use the same regulator function $R$ in Eq. (\ref{Action_Regularized}). The background covariant derivatives of the quantum gauge superfield $v$ are given by the expressions\footnote{Note in our conventions, $v^+ = e^{-2\bm{V}} v\, e^{2\bm{V}}$.}

\begin{equation}
\bm{\nabla}_a v \equiv D_a v;\qquad \bm{\xbar\nabla}_{\dot a} v \equiv e^{2\bm{V}} \bar D_{\dot a} \big(e^{-2\bm{V}} v\, e^{2\bm{V}}\big) e^{-2\bm{V}},
\end{equation}

\noindent
where $\bm{V}$ is the background gauge superfield. The corresponding actions for the Faddeev--Popov and Nielsen--Kallosh ghosts can be found in, e.g., \cite{Aleshin:2016yvj,Kazantsev:2017fdc}.

It is important that the replacement of the action (\ref{Action}) with the regularized action (\ref{Action_Regularized}) does not remove the one-loop divergences. For regularizing these residual divergences it is possible to insert the Pauli--Villars determinants into the generating functional \cite{Slavnov:1977zf}. The details of the corresponding construction in the supersymmetric case are not presented here and can again be found in \cite{Aleshin:2016yvj,Kazantsev:2017fdc}.

We will be interested in the quantum corrections to the gauge coupling constant, which can be found by calculating the two-point Green function of the background gauge superfield. Due to the manifest background gauge invariance of the effective action, it is transversal,

\begin{equation}\label{D_Inverse_Definition}
\Gamma^{(2)}_{\bm{V}} = - \frac{1}{8\pi} \mbox{tr} \int \frac{d^4p}{(2\pi)^4}\, d^4\theta\, \bm{V}(-p,\theta)\,\partial^2\Pi_{1/2} \bm{V}(p,\theta)\, d^{-1}\big(\alpha_0,\lambda_0,p,\Lambda\big),
\end{equation}

\noindent
where

\begin{equation}
\partial^2\Pi_{1/2} \equiv - \frac{1}{8} D^a \bar D^2 D_a = - \frac{1}{8} \bar D^{\dot a} D^2 \bar D_{\dot a}
\end{equation}

\noindent
is a supersymmetric transversal projector operator. Note that the function $d^{-1}$, defined by Eq. (\ref{D_Inverse_Definition}), is normalized in such a way that in the tree approximation in the limit $\Lambda\to\infty$ it is equal to $\alpha_0^{-1}$. Due to the presence of the operator $\partial^2\Pi_{1/2}$ in Eq. (\ref{D_Inverse_Definition}), the (non-regularized) integrals that determine the function $d^{-1}$ have the degree of divergence $\omega = n_\lambda$ (instead of $2+n_\lambda$ following from the expression (\ref{Degree_Of_Divegence})).

In what follows, we will discuss the relation between the leading divergences to the function $d^{-1}$ and the leading divergences to the two-point Green function of the chiral matter superfields, which can be presented in the form

\begin{equation}\label{G_Definition}
\Delta\Gamma^{(2)}_\phi = \frac{1}{4} \int \frac{d^4p}{(2\pi)^4}\, d^4\theta\, \phi^{*i}(-p,\theta)\,\phi_j(p,\theta)\, G_i{}^j\big(\alpha_0,\lambda_0,p,\Lambda\big).
\end{equation}

\noindent
In the tree approximation the function $G_i{}^j$ is evidently equal to $\delta_i^j$ in the limit $\Lambda\to \infty$, while the higher order corrections can be found by calculating supergraphs with two external lines corresponding to the superfields $\phi^*$ and $\phi$.

\section{The lowest quantum corrections proportional to $\lambda_0^*\lambda_0$}
\hspace*{\parindent}\label{Section_Quantum_Corrections}

\begin{figure}[h]
\begin{picture}(0,2.5)
\put(1.5,0){\includegraphics[scale=0.21]{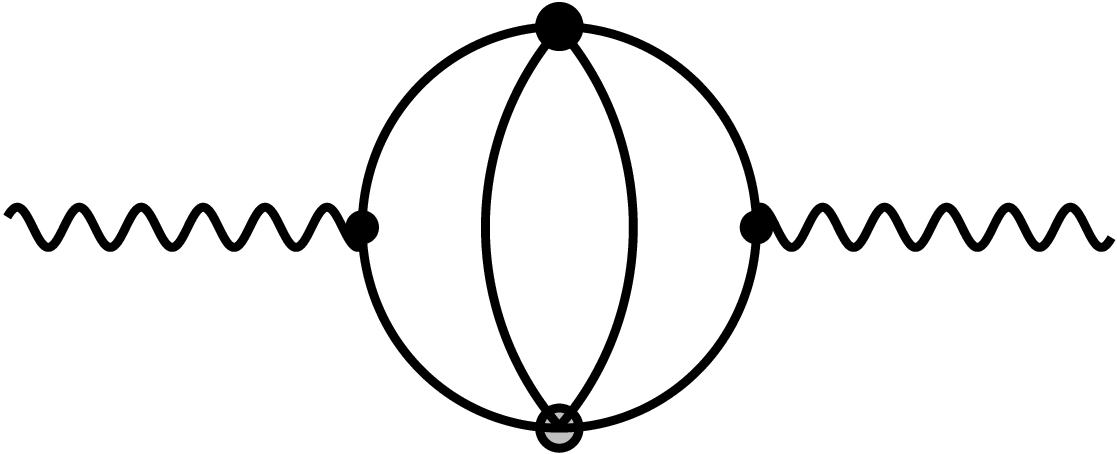}}
\put(1.5,1.6){(1)}
\put(6.7,0){\includegraphics[scale=0.18]{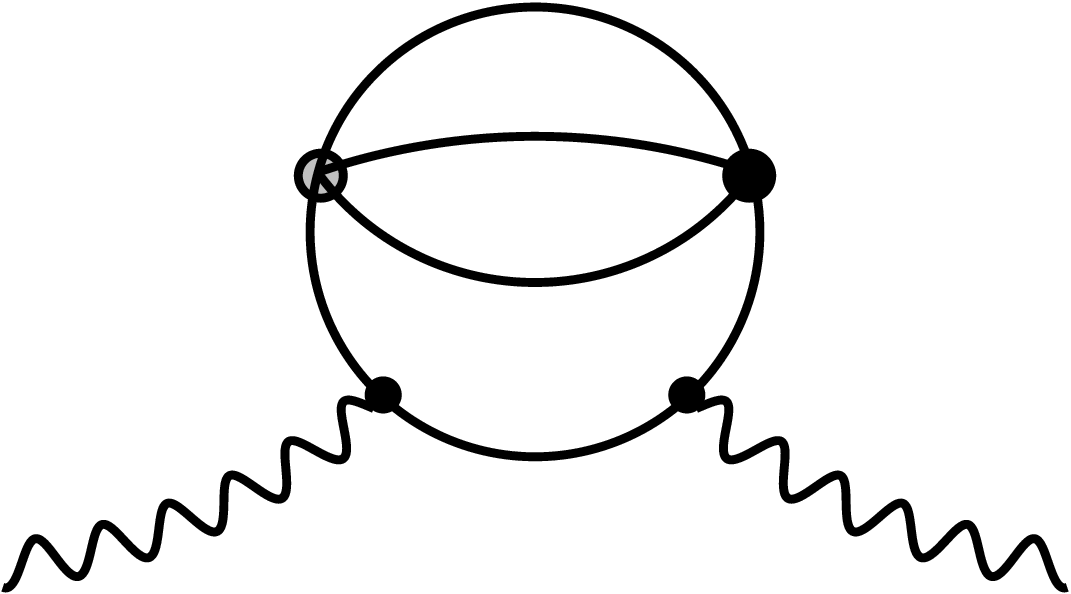}}
\put(6.5,1.6){(2)}
\put(11.8,0){\includegraphics[scale=0.18]{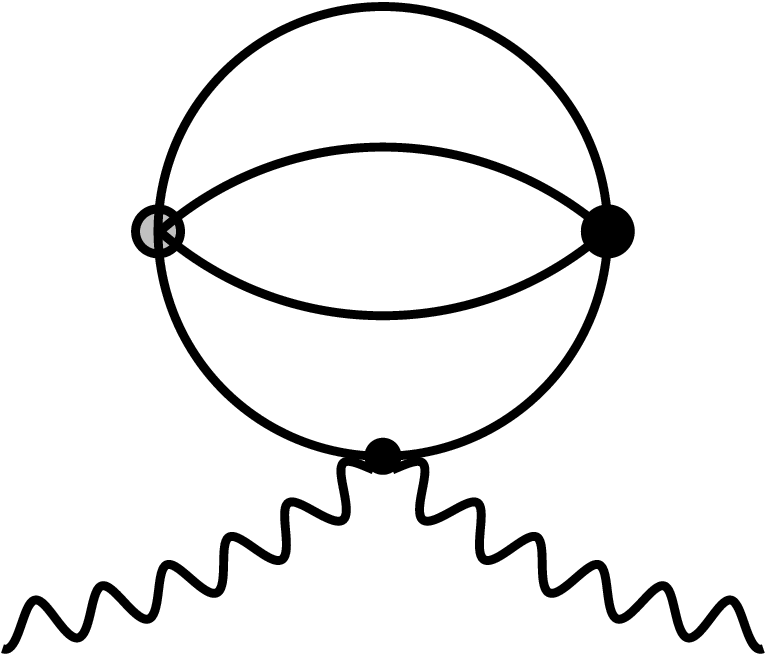}}
\put(11.1,1.6){(3)}
\end{picture}
\caption{The lowest-order supergraphs proportional to $\lambda_0^*\lambda_0$ giving a quadratically divergent contribution to the two-point Green function of the background gauge superfield}
\label{Figure_Beta}
\end{figure}
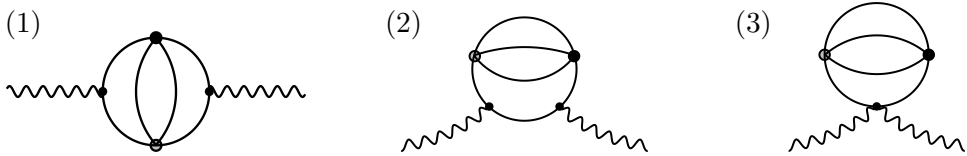

Let us calculate the lowest loop correction to the two-point Green function of the background gauge superfield proportional to $\lambda_0^*\lambda_0$. (Here we will be interested only in the contributions proportional to the quartic Yukawa couplings and will omit all other terms.) The first correction of the considered type appears in the three-loop approximation and is given by three superdiagrams presented in Fig.~\ref{Figure_Beta}. In these diagrams the external lines correspond to the background superfield $\bm{V}$. The small disks denote the gauge-matter vertices, while the grey and black large disks stand for the vertices with quartic Yukawa couplings $\lambda^*_0$ and $\lambda_0$, respectively. The expressions for them can be constructed in the standard way with the help of the superfield Feynman rules. Certainly, doing this, it is necessary to take into account terms containing higher derivatives in the regularized action (\ref{Action_Regularized}), which significantly complicate the expressions for the vertices containing the gauge superfields. However, the calculations can be made using the standard technique described, e.g., in \cite{West:1990tg} in detail. The results for all supergraphs presented in Fig.~\ref{Figure_Beta} are collected in Appendix \ref{Appendix_Gauge_Superdiagrams}.

According to Eq. (\ref{Degree_Of_Divegence}), all these supergraphs formally have the degree of divergence $\omega=4$. However, the supergraphs contributing to the two-point Green function of the background gauge superfields contain two parts. One of them is proportional to $\mbox{tr}\, \bm{V}^2$, while the other includes $\mbox{tr}\,\bm{V}\partial^2\Pi_{1/2}\bm{V}$. Due to the Slavnov--Taylor identities \cite{Taylor:1971ff,Slavnov:1972fg} (which for the background gauge transformations are reduced to the manifest background gauge invariance of the effective action), the noninvariant terms proportional to $\mbox{tr}\,\bm{V}^2$ cancel each other. This cancellation can be considered as a nontrivial correctness test and has been verified in Appendix \ref{Appendix_Gauge_Superdiagrams}. In the remaining terms four spinor derivatives inside $\partial^2\Pi_{1/2}$ act on external gauge lines thereby reducing the degree of divergence by 2. This implies that the contribution of the diagrams presented in Fig.~\ref{Figure_Beta} to the function $d^{-1}$ defined by Eq. (\ref{D_Inverse_Definition}) is in fact quadratically divergent. The leading divergences can be calculated in the limit of the vanishing external momentum. In this limit (after the Wick rotation), the contribution to the function $d^{-1}$ coming from the diagrams shown in Fig.~\ref{Figure_Beta} can be written in the form of an integral of double total derivatives with respect to loop momenta,

\begin{eqnarray}\label{Double_Derivative}
&& \Delta d^{-1}\Big|_{p\to 0} = - \frac{2\pi}{3r} \lambda^*_{0ijkn} \lambda_0^{ijkm} C(R)_m{}^n \int \frac{d^4Q}{(2\pi)^4}\, \frac{d^4K}{(2\pi)^4}\,  \frac{d^4L}{(2\pi)^4}\, \nonumber\\
&&\qquad\qquad\qquad\qquad\qquad\quad \times \frac{\partial^2}{\partial Q_\mu^2} \bigg(\frac{1}{Q^2 F_Q K^2 F_K L^2 F_L (Q+K+L)^2 F_{Q+K+L}}\bigg).\qquad
\end{eqnarray}

\noindent
In our conventions, the capital letters denote Euclidean momenta, and we use the notations $F_Q\equiv F(Q^2/\Lambda^2)$ etc. It is important that writing the integral (\ref{Double_Derivative}) we assume that small vicinities of singular points $Q_\mu=0$ and $(Q+K+L)_\mu=0$ are excluded from the integration region.

With the help of the divergence theorem, the integral of the double total derivative can be rewritten as integrals over the infinitely large sphere $S^3_\infty$ and the spheres $S^3_\varepsilon$ of an infinitely small radius $\varepsilon$ surrounding the singularities. The integral over $S^3_\infty$ vanishes due to the presence of the higher derivative regulators, while the integrals over $S^3_\varepsilon$ produce nontrivial results. For instance, if $f(Q^2/\Lambda^2)$ is a nonsingular function rapidly decreasing at infinity, then

\begin{equation}
\int \frac{d^4Q}{(2\pi)^4} \frac{\partial^2}{\partial Q_\mu^2} \bigg(\frac{f(Q^2/\Lambda^2)}{Q^2}\bigg) = \frac{1}{16\pi^4} \oint\limits_{S^3_\varepsilon} dS^{(Q)}_\mu \frac{\partial}{\partial Q_\mu} \bigg( \frac{f(Q^2/\Lambda^2)}{Q^2}\bigg) = \frac{1}{4\pi^2} f(0).
\end{equation}

\noindent
Using the same similar technique for calculating the integral in Eq. (\ref{Double_Derivative}) (and taking into account that there are two singular points in this case) we obtain the result for the considered contribution to the two-point Green function of the background gauge superfield, which contains only two loop integrations,

\begin{equation}\label{D_Inverse_Result}
\Delta d^{-1}\Big|_{p\to 0} = - \frac{1}{3\pi r} \lambda^*_{0ijkn} \lambda_0^{ijkm} C(R)_m{}^n \int \frac{d^4K}{(2\pi)^4}\,  \frac{d^4L}{(2\pi)^4}\, \frac{1}{K^2 F_K L^2 F_L (K+L)^2 F_{K+L}}.
\end{equation}

\noindent
In particular, we see that the loop integral in this expression is really quadratically divergent in agreement with the above discussion.

\begin{figure}[h]
\begin{picture}(0,2)
\put(6.5,0){\includegraphics[scale=0.2]{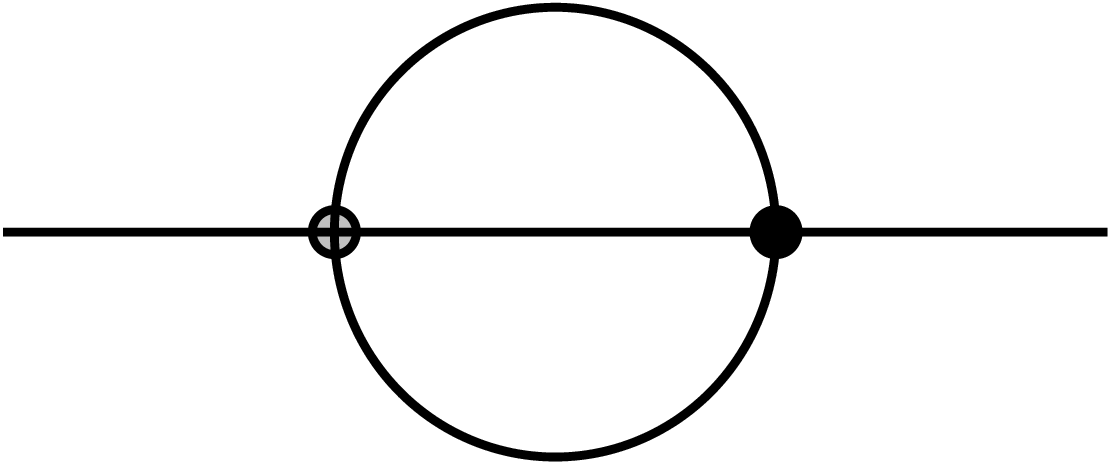}}
\put(6.5,1.03){$\phi^*$} \put(9.9,1.03){$\phi$}
\end{picture}
\caption{The lowest-order superdiagram proportional to $\lambda_0^*\lambda_0$ giving quadratically divergent contribution to the two-point Green function of the chiral matter superfields}
\label{Figure_Gamma}
\end{figure}

For renormalizable ${\cal N}=1$ supersymmetric theories the double total derivative structure of loop integrals leads to the relation between the divergent contributions in the gauge and matter sectors expressed by the NSVZ equation (\ref{NSVZ_Equation}). That is why we can expect here that the expression (\ref{D_Inverse_Result}) is related to the lowest contribution proportional to $\lambda_0^*\lambda_0$ in the two-point Green function of the matter superfields. This contribution is determined by the superdiagram presented in Fig.~\ref{Figure_Gamma}. The straightforward calculation gives the result for the corresponding part of the effective action,

\begin{eqnarray}
&& \Delta\Gamma^{(2)}_\phi = \frac{1}{6}\lambda^*_{0imnp} \lambda_0^{jmnp} \int \frac{d^4p}{(2\pi)^4} d^4\theta\, \phi^{*i}(-p,\theta)\,\phi_j(p,\theta) \nonumber\\
&&\qquad\qquad\qquad\qquad\qquad\qquad \times \int \frac{d^4k}{(2\pi)^4}\,  \frac{d^4l}{(2\pi)^4}\, \frac{1}{k^2 F_k l^2 F_l (k+l+p)^2 F_{k+l+p}},\qquad
\end{eqnarray}

\noindent
where $F_k\equiv F(-k^2/\Lambda^2)$ etc. Next, it is necessary to extract from this equation a part of the function $G_i{}^j$ defined by Eq. (\ref{G_Definition}) and make the Wick rotation. After that, in the limit of the vanishing external momentum $p\to 0$ we obtain the expression

\begin{equation}\label{Gamma_Result}
\Delta G_i{}^j\Big|_{p\to 0} = \frac{2}{3}\lambda^*_{0imnp} \lambda_0^{jmnp} \int \frac{d^4K}{(2\pi)^4}\,  \frac{d^4L}{(2\pi)^4}\, \frac{1}{K^2 F_K L^2 F_L (K+L)^2 F_{K+L}},
\end{equation}

\noindent
where all integrations are made over the Euclidean loop momenta. Comparing this result with Eq. (\ref{D_Inverse_Result}), we conclude that the quadratically divergent quantum corrections to the gauge coupling and to the kinetic term of the chiral matter superfields produced by the supergraphs in Figs. \ref{Figure_Beta} and \ref{Figure_Gamma} are related by the NSVZ-like equation

\begin{equation}\label{Relation_Between_Green_Functions}
\Delta d^{-1}\Big|_{p\to 0} = - \frac{1}{2\pi r} C(R)_i{}^j \Delta G_j{}^i\Big|_{p\to 0},
\end{equation}

\noindent
which is valid at the level of loop integrals. In fact, it is not easy to calculate the integral in Eq. (\ref{Gamma_Result}) even for the simplest choice of the regulator, $F(x)=1+x$. In this case it can be related to the quadratically divergent massive sunrise graph, see \cite{Fujikawa:2011zf}. Wide literature concerning this issue can be found, e.g., in \cite{Kalmykov:2016lxx}. With the help of the results obtained in \cite{Davydychev:1992mt}, the leading quadratic divergence given by the expression (\ref{Gamma_Result}) for the regulator function $F(x)=1+x$ has been calculated in Appendix \ref{Appendix_Integral}.

Note that for renormalizable theories, the relations similar to (\ref{Relation_Between_Green_Functions}) produce the all-loop NSVZ equation for RGFs (\ref{RGFs_Bare}) defined in terms of the bare couplings, which are related to the Green functions by the equations

\begin{equation}
\frac{\beta_0(\alpha_0,\lambda_0)}{\alpha_0^2} \equiv \frac{d}{d\ln\Lambda} \Big(d^{-1} - \alpha_0^{-1}\Big)\bigg|_{\alpha,\lambda=\mbox{\scriptsize const};\ p\to 0};\qquad
(\gamma_{0})_i{}^j(\alpha_0,\lambda_0) \equiv \frac{d\ln G_i{}^j}{d\ln\Lambda} \bigg|_{\alpha,\lambda=\mbox{\scriptsize const};\ p\to 0},
\end{equation}

\noindent
where the derivatives are calculated at fixed values of the renormalized couplings. The NSVZ equation is then satisfied in all orders in the case of using the higher covariant derivative regularization and (for a fixed regularization) does not depend on a particular renormalization procedure, because the RGFs (\ref{RGFs_Bare}) are independent of it \cite{Kataev:2013eta}. In the nonrenormalizable case we obtained an analog of this statement. By analogy with the renormalizable case, it seems to be always valid with the higher covariant derivative regularization, however, it is certainly impossible to discuss renormalization prescriptions and renormalized couplings in this case.

\section{The calculation based on the vacuum supergraphs}
\hspace*{\parindent}\label{Section_Vacuum}

As we have already mentioned, the double total derivatives appear in all loop integrals which determine the $\beta$-function of ${\cal N}=1$ supersymmetric theories regularized by higher covariant derivatives \cite{Stepanyantz:2019ihw}. The corresponding superdiagrams contain two external lines of the background gauge superfield, as the supergraphs presented in Fig.~\ref{Figure_Beta}. However, in the case of renormalizable supersymmetric theories it is possible to obtain the integrals of double total derivatives by calculating only (properly modified) vacuum supergraphs. This method was proposed in \cite{Stepanyantz:2019ihw} and was subsequently applied to various explicit multiloop calculations in Refs. \cite{Stepanyantz:2019lyo,Kuzmichev:2019ywn,Aleshin:2020gec,Aleshin:2022zln,Shirokov:2023jya}.

\begin{figure}[h]
\begin{picture}(0,2.5)
\put(7.4,0){\includegraphics[scale=0.21]{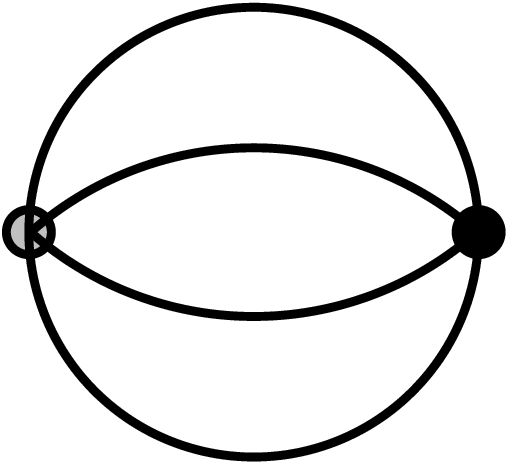}}
\end{picture}
\caption{The vacuum supergraph generating the contributions presented in Fig.~\ref{Figure_Beta}}
\label{Figure_Vacuum}
\end{figure}

The main idea of the method is that considering a vacuum supergraph we can calculate a certain contribution to the two-point Green function of the background gauge superfield $\bm{V}$. This contribution is produced by all diagrams that are obtained by adding two external $\bm{V}$-lines to the original vacuum supergraph in all possible ways. In particular, the superdiagrams presented in Fig.~\ref{Figure_Beta} are generated by a single supergraph depicted in Fig.~\ref{Figure_Vacuum}.

In application to nonrenormalizable theories, for obtaining a leading power divergent contribution to the function $d^{-1}$ corresponding to a certain vacuum supergraph with $L\ge 2$ loops the algorithm proposed in \cite{Stepanyantz:2019ihw} can be formulated as follows.

1. First, we construct the expression for a vacuum supergraph with the help of the Feynman rules and insert into an arbitrary point a factor $\theta^4 \rho$, where $\theta^4 \equiv \theta^a \theta_a \bar\theta^{\dot a}\bar\theta_{\dot a}$ and $\rho$ is a function of the space-time coordinates which slowly tends to 0 at a very large distance $X$. (Without the insertion of $\theta^4$ one will certainly obtain the vanishing result.)

2. The supergraph should be calculated according to the usual rules (see, e.g., \cite{West:1990tg}), but the terms suppressed by powers of $1/(\Lambda X)$ should be omitted. Note that the result will contain the factor

\begin{equation}
{\cal V}_4 \equiv \int d^4x \rho(x).
\end{equation}

3. Next, it is necessary to mark $L$ internal lines with independent (Euclidean) momenta $Q^\mu_i$ with $i=1,\ldots,L$, where $L$ is the number of loops in the vacuum supergraph under consideration. We denote internal indices corresponding to the beginnings and endings of these lines by $a_i$ and $b_i$, respectively, so that the product of the marked propagators will be proportional to $\prod_i \delta_{a_i}^{b_i}$.

4. In the loop integral one should change the integrand by making the replacement

\begin{equation}\label{Replacement}
\prod\limits_i \delta_{a_i}^{b_i} \to \sum\limits_{m,n=1}^L \prod\limits_{i\ne m,n} \delta_{a_i}^{b_i} (T^A)_{a_m}{}^{b_m} (T^A)_{a_n}{}^{b_n} \frac{\partial^2}{\partial Q_m^\mu\,\partial Q_n^\mu}.
\end{equation}

\noindent
(Certainly, after this replacement we obtain an integral of double total derivatives with respect to the loop momenta.)

5. Finally the contribution\footnote{Let us recall that we calculate the sum of superdiagrams obtained from the original vacuum supergraph by attaching two external $\bm{V}$-lines in all possible ways.} to the function $d^{-1}$ in the limit of the vanishing external momentum ($p\to 0$) will be obtained by multiplying the result by the factor

\begin{equation}\label{Factor}
-\frac{2\pi}{r{\cal V}_4},
\end{equation}

\noindent
where $r$ is the dimension of the gauge group.

The derivation of this algorithm in \cite{Stepanyantz:2019ihw} is rather complicated. Although it has been made under the assumption of renormalizability, the proof also seems to be valid for theories with nonrenormalizable superpotential if the leading power divergent terms in the function $d^{-1}$ are calculated in the limit $p\to 0$.

The calculation of the vacuum supergraph presented in Fig.~\ref{Figure_Vacuum} modified according to the instruction 1 of the above algorithm gave the result

\begin{equation}\label{Original_Vaccum_Supergraph_Expression}
\frac{1}{6} {\cal V}_4\, \lambda_0^{ijkl} \lambda^*_{0mnpl} \int \frac{d^4Q}{(2\pi)^4}\,\frac{d^4K}{(2\pi)^4}\,\frac{d^4L}{(2\pi)^4}\, \delta_i^m \delta_j^n \delta_k^p\,\frac{1}{Q^2 F_Q K^2 F_K L^2 F_L (Q+K+L)^2 F_{Q+K+L}},
\end{equation}

\noindent
where we have extracted the product of $\delta$-symbols corresponding to the propagators with the independent Euclidean momenta $Q_\mu$, $K_\mu$, and $L_\mu$. (The remaining loop momentum $-(Q+K+L)_\mu$ can be expressed in terms of them.) The momenta $Q_\mu$, $K_\mu$, and $L_\mu$ appear in the expression (\ref{Original_Vaccum_Supergraph_Expression}) symmetrically. This implies that, after performing the replacement (\ref{Replacement}) and multiplying the result by the factor (\ref{Factor}), the product of $\delta$-symbols inside the integrand of Eq. (\ref{Original_Vaccum_Supergraph_Expression}) should be formally converted into the operator

\begin{equation}\label{Replacement_Operator}
\delta_i^m \delta_j^n \delta_k^p\, \to\, -\frac{2\pi}{r{\cal V}_4} \bigg\{ 3 C(R)_i{}^m \delta_j^n \delta_k^p \frac{\partial^2}{\partial Q_\mu^2} + 6 (T^A)_i{}^m (T^A)_j{}^n \delta_k^p \frac{\partial^2}{\partial Q_\mu\, \partial K_\mu} \bigg\}.
\end{equation}

\noindent
Next, it is possible to simplify the resulting expression with the help of the identity

\begin{equation}\label{Lambda_Identity}
\lambda_0^{ijkl} \lambda^*_{0mnkl} (T^A)_i{}^m (T^A)_j{}^n = - \frac{1}{3} \lambda^*_{0ijkn} \lambda_0^{ijkm} C(R)_m{}^n,
\end{equation}

\noindent
which follows from Eq. (\ref{Yukawa_Invariance}), and the identity

\begin{eqnarray}\label{Integral_Identity}
&& \int \frac{d^4Q}{(2\pi)^4}\,\frac{d^4K}{(2\pi)^4}\,\frac{d^4L}{(2\pi)^4}\,\frac{\partial^2}{\partial Q_\mu\, \partial K_\mu}\bigg(\frac{1}{Q^2 F_Q K^2 F_K L^2 F_L (Q+K+L)^2 F_{Q+K+L}}\bigg)\nonumber\\
&&\qquad\qquad
= \frac{1}{2} \int \frac{d^4Q}{(2\pi)^4}\,\frac{d^4K}{(2\pi)^4}\,\frac{d^4L}{(2\pi)^4}\,\frac{\partial^2}{\partial Q_\mu^2}\bigg(\frac{1}{Q^2 F_Q K^2 F_K L^2 F_L (Q+K+L)^2 F_{Q+K+L}}\bigg),
\qquad
\end{eqnarray}

\noindent
which can be verified by a straightforward calculation. Making the replacement (\ref{Replacement_Operator}) in Eq. (\ref{Original_Vaccum_Supergraph_Expression}) and taking into account the identities (\ref{Lambda_Identity}) and (\ref{Integral_Identity}), we obtain that the contribution to the function $d^{-1}$ coming from the supergraphs presented in Fig.~\ref{Figure_Beta} should be given by the expression

\begin{eqnarray}
&& \Delta d^{-1}\Big|_{p\to 0} = - \frac{2\pi}{3r} \lambda^*_{0ijkn} \lambda_0^{ijkm} C(R)_m{}^n \int \frac{d^4Q}{(2\pi)^4}\, \frac{d^4K}{(2\pi)^4}\,  \frac{d^4L}{(2\pi)^4}\, \nonumber\\
&&\qquad\qquad\qquad\qquad\qquad\quad \times \frac{\partial^2}{\partial Q_\mu^2} \bigg(\frac{1}{Q^2 F_Q K^2 F_K L^2 F_L (Q+K+L)^2 F_{Q+K+L}}\bigg).\qquad
\end{eqnarray}

\noindent
We see that this result coincides exactly with the expression (\ref{Double_Derivative}) obtained earlier by the straightforward calculation of superdiagrams with two external $\bm{V}$-lines. Thus, even for the nonrenormalizable ${\cal N}=1$ supersymmetric theories it seems possible to calculate leading power divergent contributions to the function $d^{-1}$ considering only (specially modified) vacuum supergraphs with the help of the algorithm described above.

\begin{figure}[h]
\begin{picture}(0,7.2)
\put(0.2,5.1){\includegraphics[scale=0.18]{beta_nonren1.eps}}
\put(0.9,0){\includegraphics[scale=0.162]{beta_nonren3.eps}}
\put(0.5,2.5){\includegraphics[scale=0.162]{beta_nonren2.eps}}
\put(3.8,3.0){${\displaystyle
\left.
\begin{array}{c}
\vphantom{1}\\
\vphantom{1}\\
\vphantom{1}\\
\vphantom{1}\\
\vphantom{1}\\
\vphantom{1}\\
\vphantom{1}\\
\vphantom{1}\\
\vphantom{1}\\
\vphantom{1}\\
\vphantom{1}\\
\vphantom{1}\\
\vphantom{1}
\end{array}
\right\}
}$}
\put(5.2,2.88){\includegraphics[scale=0.22]{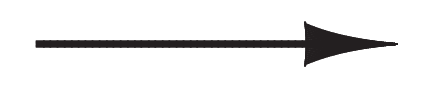}}
\put(7.7,2.4){\includegraphics[scale=0.18]{vacuum_nonren.eps}}
\put(9.7,2.88){\includegraphics[scale=0.22]{arrow_new.eps}}
\put(12,2.4){\includegraphics[scale=0.18]{matter_nonren.eps}}
\end{picture}
\caption{Graphical illustration of the equation relating divergences of different types in the theory under consideration}
\label{Figure_General}
\end{figure}

The relation between the divergent contributions of various structures can be graphically interpreted as shown in Fig.~\ref{Figure_General}. The divergent quantum corrections to the gauge coupling come from the superdiagrams with two external gauge lines presented on the left. They are obtained from the corresponding vacuum supergraph (presented in the center of Fig.~\ref{Figure_General}) by attaching two external legs in all possible ways. As has been demonstrated, (in the supersymmetric case, with the higher covariant derivative regularization) the sum of the left superdiagrams can be obtained by calculating this vacuum supergraph after some special modifications. The result is always given by an integral of double total derivatives with respect to the loop momenta. These double total derivatives effectively cut an internal line and produce a quantum correction to the kinetic term of the chiral matter superfields coming from the superdiagram depicted in Fig.~\ref{Figure_General} on the right. Consequently, quantum corrections to the gauge coupling constant and to the kinetic term of the chiral matter superfields turn out to be related by Eq. (\ref{Relation_Between_Green_Functions}) (at least for the supergraphs under consideration). This picture is quite analogous to the similar picture in the renormalizable case, see, e.g., \cite{Smilga:2004zr,Shakhmanov:2017soc,Kazantsev:2018nbl}. Therefore, we see that a relation analogous to the NSVZ equation (in the form proposed in \cite{Stepanyantz:2016gtk}, where the $\beta$-function in a certain loop is related to the anomalous dimensions of {\it all} quantum superfields in the {\it previous} loop), seems to be valid for leading power divergences in nonrenormalizable ${\cal N}=1$ supersymmetric theories.

\section*{Conclusion}
\hspace*{\parindent}

In this paper we demonstrated that some features of quantum corrections in ${\cal N}=1$ supersymmetric theories regularized by higher covariant derivatives remain valid if the superpotential contains terms of more than the third degree in the chiral matter superfields. The presence of such terms breaks renormalizability because the corresponding couplings have the dimension of mass in negative powers. Nevertheless, they are often used for constructing theories describing physics beyond the Standard Model. In principle, they can appear effectively after integrating out certain very massive superfields. If the corresponding masses are about the Planck mass, then the  couplings in the nonrenormalizable terms produced in this way are suppressed by its inverse powers. Therefore, although the nonrenormalizable theories under consideration contain power divergences, the corresponding quantum corrections are not large if we assume that all divergences cancel each other in a certain ``good'' theory above the Planck scale. In particular, this may occur due to the presence of some higher derivative terms similar to those that were used in this paper for introducing the regularization.

If a nonrenormalizable ${\cal N}=1$ supersymmetric theory with the quartic superpotential is regularized by higher derivatives, then, as shown in this paper in the lowest nontrivial approximation, the leading power divergent contribution to the two-point Green function of the background gauge superfield is related to the corresponding contribution to the two-point Green function of the chiral matter superfields. This presumably implies that, for nonrenormalizable ${\cal N}=1$ supersymmetric theories, there is a certain relation between leading divergent contributions to the gauge coupling(s) and to the Green functions of various quantum superfields. This relation (given by Eq. (\ref{Relation_Between_Green_Functions}) in the case under consideration) is analogous to the exact NSVZ equation for renormalizable theories (written in the form of the relation between the $\beta$-function and the anomalous dimensions of the quantum gauge superfield, Faddeev--Popov ghosts, and chiral matter superfields, see \cite{Stepanyantz:2016gtk} for details). In particular, exactly as for the NSVZ equation, it appears because the integrals that determine the leading divergent contributions in the gauge part of the action are given by integrals of double total derivatives with respect to the loop momenta and, exactly as in the renormalizable case, can be constructed by calculating properly modified vacuum supergraphs. The double total derivative structure of loop integrals allows for calculating one of them thereby reducing the number of integrations by 1. Moreover, just as in the renormalizable case, taking the integral of a double total derivative can be graphically interpreted as cutting an internal line. In the case under consideration, this cutting produces a certain contribution to the two-point Green function of the matter superfields, as illustrated in Fig.~\ref{Figure_General}. Thus, for nonrenormalizable ${\cal N}=1$ supersymmetric theories it seems possible to construct an analog of the NSVZ relation. In this case it appears as a certain equation relating various Green functions of the (regularized) theory. Of course, the very existence of Eq. (\ref{Relation_Between_Green_Functions}) seems rather surprising, because the Green functions of the background gauge superfield and of the chiral matter superfields correspond to quite different contributions to the effective action. Therefore, it would be interesting to find out whether it is possible to derive such relations from a certain underlying principle, e.g., with the help of arguments based on symmetries and anomalies. Another interesting problem is to reveal whether the NSVZ-like equations for the nonrenormalizable theories  indicates the presence of hidden dualities \cite{Seiberg:1994pq}. We hope to continue investigating all these issues in more detail in future research.

\section*{Acknowledgements}
\hspace*{\parindent}

K.S. would like to express his gratitude to M.~Yu.~Kalmykov for explaining some issues related to the sunrise integrals and pointing out some important references. K.S. is also very grateful to I.~E.~Shirokov for providing some information concerning the loop integrals.

\appendix

\section*{Appendix}

\section{Expressions for the superdiagrams presented in Fig.~\ref{Figure_Beta}}
\hspace*{\parindent}\label{Appendix_Gauge_Superdiagrams}

For completeness, here we present the explicit expressions for the superdiagrams depicted in Fig.~\ref{Figure_Beta}. They are written in Minkowski space, where the momenta are denoted by lowercase letters (unlike the Euclidean momenta, which are denoted by capital letters).

\begin{eqnarray}
&&\hspace*{-5mm} (1) = \frac{2i}{r}\lambda_0^{ijkl} \lambda^*_{0mnkl} (T^A)_i{}^m (T^A)_j{}^n\, \mbox{tr}\int \frac{d^4p}{(2\pi)^4}\,d^4\theta\,
\int\frac{d^4q}{(2\pi)^4}\,\frac{d^4k}{(2\pi)^4}\,\frac{d^4l}{(2\pi)^4} \bigg\{\bm{V}(-p,\theta) \bm{V}(p,\theta) \nonumber\\
&&\hspace*{-5mm} \times \frac{q^2 F_q F_{q+k+l+p}}{(q+k+l)^2} + \bm{V}(-p,\theta) \partial^2\Pi_{1/2} \bm{V}(p,\theta) \bigg(\frac{(q+p)^2 F_{q+p} - q^2 F_q}{(q+p)^2-q^2}\bigg) \nonumber\\
&&\hspace*{-5mm} \times \bigg[- \frac{(k+l)^2}{2(q+k+l)^2 (q+k+l+p)^2}\bigg(\frac{(q+k+l+p)^2 F_{q+k+l+p}-(q+k+l)^2 F_{q+k+l}}{(q+k+l+p)^2-(q+k+l)^2}\bigg) \nonumber\\
&&\hspace*{-5mm} + \frac{F_{q+k+l+p}- F_{q+k+l}}{(q+k+l+p)^2-(q+k+l)^2}\bigg]\bigg\}\frac{1}{q^2 F_q (q+p)^2 F_{q+p} k^2 F_k l^2 F_l F_{q+k+l} F_{q+k+l+p}};\\
&& \vphantom{\Big(}\nonumber\\
&&\hspace*{-5mm} (2) = \frac{2i}{3r}\lambda^*_{0ijkn} \lambda_0^{ijkm} C(R)_m{}^n\,\mbox{tr} \int \frac{d^4p}{(2\pi)^4}\,d^4\theta\,
\int\frac{d^4q}{(2\pi)^4}\,\frac{d^4k}{(2\pi)^4}\,\frac{d^4l}{(2\pi)^4} \bigg\{\bm{V}(-p,\theta) \bm{V}(p,\theta)\Big(q^2 F_q^2\nonumber\\
&&\hspace*{-5mm} + (q+p)^2 F_{q+p}^2\Big) - \bm{V}(-p,\theta) \partial^2\Pi_{1/2} \bm{V}(p,\theta)\bigg[\bigg(\frac{(q+p)^2 F_{q+p} - q^2 F_q}{(q+p)^2-q^2}\bigg)^2
+ \bigg(\frac{F_{q+p}-F_q}{(q+p)^2-q^2}\bigg)^2  \nonumber\\
&&\hspace*{-5mm} \times (q+p)^2 q^2 \bigg] \bigg\} \frac{1}{(q+p)^2 F_{q+p} q^2 F_q^2 k^2 F_k l^2 F_l (q+k+l)^2 F_{q+k+l}};\\
&& \vphantom{\Big(}\nonumber\\
&&\hspace*{-5mm} (3) = \frac{2i}{3r}\lambda^*_{0ijkn} \lambda_0^{ijkm} C(R)_m{}^n\,\mbox{tr} \int \frac{d^4p}{(2\pi)^4}\,d^4\theta\,
\int\frac{d^4q}{(2\pi)^4}\,\frac{d^4k}{(2\pi)^4}\,\frac{d^4l}{(2\pi)^4} \bigg\{-\bm{V}(-p,\theta) \bm{V}(p,\theta) F_{q+p}\nonumber\\
&&\hspace*{-5mm} + \bm{V}(-p,\theta) \partial^2\Pi_{1/2} \bm{V}(p,\theta)\bigg[\frac{(q+p)^2+q^2}{((q+p)^2-q^2)^2}(F_{q+p}-F_q)
+ \frac{2q^2 F'_q}{\Lambda^2((q+p)^2-q^2)} \bigg]\bigg\}\frac{1}{q^2 F_q^2 k^2 F_k}\nonumber\\
&&\hspace*{-5mm} \times \frac{1}{l^2 F_l (q+k+l)^2 F_{q+k+l}}.
\end{eqnarray}

\noindent
Here we use the notations $F_q\equiv F(-q^2/\Lambda^2)$, $F'_q = F'(-q^2/\Lambda^2)$ with the prime denoting the derivative with respect to the argument $-q^2/\Lambda^2$.

As a correctness test, (with the help of the identity (\ref{Lambda_Identity})) it is possible to verify that all terms proportional to $\mbox{tr}(\bm{V}(-p,\theta) \bm{V}(p,\theta))$ cancel each other,

\begin{eqnarray}\label{Green_Function}
&&\hspace*{-5mm} \Delta\Gamma^{(2)}_{\bm{V}} = (1) + (2) + (3) = \frac{2i}{3r}\lambda^*_{0ijkn} \lambda_0^{ijkm} C(R)_m{}^n\,\mbox{tr} \int \frac{d^4p}{(2\pi)^4}\,d^4\theta\,\bm{V}(-p,\theta)\partial^2\Pi_{1/2} \bm{V}(p,\theta)\nonumber\\
&&\hspace*{-5mm} \times \int\frac{d^4q}{(2\pi)^4}\,\frac{d^4k}{(2\pi)^4}\,\frac{d^4l}{(2\pi)^4} \frac{1}{q^2 F_q (q+p)^2 F_{q+p} k^2 F_k l^2 F_l F_{q+k+l}}\bigg\{\frac{1}{F_{q+k+l+p}}\bigg(\frac{(q+p)^2 F_{q+p} - q^2 F_q}{(q+p)^2-q^2}\bigg)\nonumber\\
&&\hspace*{-5mm} \times \bigg[\frac{(k+l)^2}{2(q+k+l)^2 (q+k+l+p)^2}\bigg(\frac{(q+k+l+p)^2 F_{q+k+l+p}-(q+k+l)^2 F_{q+k+l}}{(q+k+l+p)^2-(q+k+l)^2}\bigg) \nonumber\\
&&\hspace*{-5mm} - \frac{F_{q+k+l+p}- F_{q+k+l}}{(q+k+l+p)^2-(q+k+l)^2} \bigg] + \frac{1}{(q+k+l)^2} \bigg[-F_{q+p} + \frac{q^2((q+p)^2 + q^2)}{((q+p)^2-q^2)^2}(F_{q+p}-F_q)
\nonumber\\
&&\hspace*{-5mm}
+ \frac{2q^2 (q+p)^2 F'_q F_{q+p}}{\Lambda^2((q+p)^2-q^2)F_q} \bigg]\bigg\}.
\end{eqnarray}

\noindent
Certainly, this is needed for the Green function in question to be transverse as required by the manifest background gauge invariance of the effective action.

Comparing the expression (\ref{Green_Function}) with Eq. (\ref{D_Inverse_Definition}), we obtain the corresponding contribution to the function $d^{-1}$. After the Wick rotation, in the limit of vanishing external momentum, it can be written in the form

\begin{eqnarray}
&& \Delta d^{-1}\Big|_{p\to 0} = -\frac{16\pi}{3r} \lambda^*_{0ijkn} \lambda_0^{ijkm} C(R)_m{}^n \int \frac{d^4Q}{(2\pi)^4}\, \frac{d^4K}{(2\pi)^4}\,  \frac{d^4L}{(2\pi)^4}\, \frac{1}{Q^2 F_Q K^2 F_K L^2 F_L}
\nonumber\\
&& \times \frac{1}{(Q+K+L)^2 F_{Q+K+L}}\bigg[-\frac{Q^2 F''_Q}{\Lambda^4 F_Q} + \frac{2 Q^2 (F'_Q)^2}{\Lambda^4 F_Q^2} + Q_\mu(Q+K+L)_\mu\bigg(\frac{F_Q'}{\Lambda^2 F_Q} + \frac{1}{Q^2}\bigg)\qquad\nonumber\\
&& \times \bigg(\frac{F'_{Q+K+L}}{\Lambda^2 F_{Q+K+L}} + \frac{1}{(Q+K+L)^2}\bigg) \bigg].
\end{eqnarray}

\noindent
(Here the primes denote derivatives with respect to the arguments $Q^2/\Lambda^2$ etc.) One can check that this expression can be written in the form of an integral of double total derivatives with respect to the loop momenta and coincides exactly with Eq. (\ref{Double_Derivative}).

\section{Explicit expression for the leading quadratic divergence in the particular case $F(x)=1+x$}
\hspace*{\parindent}\label{Appendix_Integral}

The simplest choice of the regulator function is $F(x)=1+x$, for which the integral in Eq. (\ref{Gamma_Result}) takes the form

\begin{eqnarray}\label{Integral}
I = \int\frac{d^4K}{(2\pi)^4}\, \frac{d^4L}{(2\pi)^4}\, \frac{1}{K^2 \Big(1+K^2/\Lambda^2\Big) L^2 \Big(1+L^2/\Lambda^2\Big) (K+L)^2\Big(1+(K+L)^2/\Lambda^2\Big)}.
\end{eqnarray}

\noindent
This expression can be simplified with the help of the identity

\begin{equation}
\frac{1}{K^2 (1+K^2/\Lambda)} = \frac{1}{K^2} - \frac{1}{K^2+\Lambda^2}.
\end{equation}

To avoid ill-defined expressions at intermediate steps of the calculation, we will use auxiliary dimensional regularization. Certainly, the original expression (\ref{Integral}) is well defined, so that the limit $\varepsilon\equiv 4-D \to 0$ should exist. In what follows, this fact will be used for verifying the correctness of the calculation. After introducing the auxiliary regularization, the expression (\ref{Integral}) can be rewritten in the form

\begin{eqnarray}
&& I = \lim\limits_{\varepsilon\to 0} \int\frac{d^DK}{(2\pi)^D}\,\frac{d^DL}{(2\pi)^D}\, \bigg(\frac{1}{K^2 L^2 (K+L)^2} - \frac{3}{(K^2+\Lambda^2) L^2 (K+L)^2}\nonumber\\
&&\qquad\qquad  + \frac{3}{(K^2+\Lambda^2) (L^2+\Lambda^2) (K+L)^2}
- \frac{1}{(K^2+\Lambda^2) (L^2+\Lambda^2) ((K+L)^2+\Lambda^2)} \bigg).\qquad
\end{eqnarray}

\noindent
The integral of the first term evidently vanishes due to the absence of a dimensionful parameter. The integral of the second term can be calculated with the help of the standard technique described, e.g., in \cite{Ramond:1981pw},

\begin{equation}
\int\frac{d^DK}{(2\pi)^D}\,\frac{d^DL}{(2\pi)^D}\, \frac{1}{(K^2+\Lambda^2) L^2 (K+L)^2} = - \frac{(\Lambda^2)^{1-\varepsilon} }{(4\pi)^{4-\varepsilon}} \cdot \frac{\Gamma(1-\varepsilon/2)\Gamma(1+\varepsilon/2)\Gamma(1+\varepsilon)}{(1-\varepsilon/2)(1-\varepsilon)}\cdot \frac{2}{\varepsilon^2}.
\end{equation}

\noindent
Two remaining integrals can be found in \cite{Davydychev:1992mt},

\begin{eqnarray}
&&\hspace*{-5mm} \int\frac{d^DK}{(2\pi)^D}\,\frac{d^DL}{(2\pi)^D}\, \frac{1}{(K^2+\Lambda^2) (L^2+\Lambda^2) (K+L)^2} = -\frac{(\Lambda^2)^{1-\varepsilon} }{(4\pi)^{4-\varepsilon}} \cdot \frac{\Gamma^2(1+\varepsilon/2)}{(1-\varepsilon/2)(1-\varepsilon)}\cdot \frac{4}{\varepsilon^2} + O(\varepsilon);\nonumber\\
&&\vphantom{1}\\
&&\hspace*{-5mm} \int\frac{d^DK}{(2\pi)^D}\,\frac{d^DL}{(2\pi)^D}\, \frac{1}{(K^2+\Lambda^2) (L^2+\Lambda^2) ((K+L)^2+\Lambda^2)}\nonumber\\
&&\hspace*{-5mm} \qquad\qquad\qquad\qquad\quad  = \frac{(\Lambda^2)^{1-\varepsilon} }{(4\pi)^{4-\varepsilon}} \cdot \frac{\Gamma^2(1+\varepsilon/2)}{(1-\varepsilon/2)(1-\varepsilon)}\cdot \Big(-\frac{6}{\varepsilon^2} + 2\sqrt{3}\, \mbox{Cl}_2(\pi/3)\Big) + O(\varepsilon),\qquad
\end{eqnarray}

\noindent
where the Clausen's integral function is defined by the equation

\begin{equation}
\mbox{Cl}_2(\theta) \equiv - \int\limits_0^\theta d\theta\, \ln|2\sin(\theta/2)|,
\end{equation}

\noindent
and, in particular, $\mbox{Cl}_2(\pi/3)\approx 1.0149416$.

Collecting the expressions presented above, we see that all $\varepsilon$-poles disappear (as they should), and the final result takes the form

\begin{eqnarray}
&& I = \lim\limits_{\varepsilon\to 0}\, \frac{(\Lambda^2)^{1-\varepsilon} }{(4\pi)^{4-\varepsilon}} \cdot \frac{\Gamma(1+\varepsilon/2)}{(1-\varepsilon/2)(1-\varepsilon)}\bigg\{\frac{6\Gamma(1-\varepsilon/2)\Gamma(1+\varepsilon)}{\varepsilon^2} +\Gamma(1+\varepsilon/2)\bigg(- \frac{12}{\varepsilon^2} +\frac{6}{\varepsilon^2}\qquad \nonumber\\
&& - 2\sqrt{3}\, \mbox{Cl}_2(\pi/3)\bigg)+O(\varepsilon)\bigg\} = \frac{\Lambda^2 }{(4\pi)^4} \cdot \Big(\frac{\pi^2}{2} - 2\sqrt{3}\, \mbox{Cl}_2(\pi/3)\Big)\approx \frac{\Lambda^2 }{(4\pi)^4} \cdot 1.418941.
\end{eqnarray}

\end{document}